\documentstyle[sprocl,epsfig]{article}

\def\be{\begin{equation}}
\def\ee{\end{equation}}
\def\bea{\begin{eqnarray}}
\def\eea{\end{eqnarray}}
\newcommand{\gev}{{\rm Ge}\kern-1.pt{\rm V}}
\newcommand{\gevsq}{\mbox{$\mathrm{{\rm Ge}\kern-1.pt{\rm V}}^2$}}

\newcommand{\rhoz}{\mbox{$\rho^0$}}
\newcommand{\jpsi}{\mbox{$J/\psi$}}

\newcommand{\qsq}{\mbox{$Q^2$}}

\newcommand{\D}{\mbox{\rm{d}}}

\newcommand{\thetah}{\mbox{${\theta}_h$}}

\def\lsim{\mathrel{\rlap{\lower4pt\hbox{\hskip1pt$\sim$}}
    \raise2pt\hbox{$<$}}} %less than or approx. symbol
\def\gsim{\mathrel{\rlap{\lower4pt\hbox{\hskip1pt$\sim$}}
    \raise2pt\hbox{$>$}}} %greater than or approx. symbol

\begin{document}
\title{Recent Results from Investigations of Diffractive Vector Meson
Production with the ZEUS Detector at HERA}
\author{J.A.~Crittenden}
\address{Physikalisches 
Institut, Universit\"at Bonn, Nu{\ss}allee 12, 53115 Bonn, Germany\\
{\rm on behalf of the ZEUS collaboration}}
\maketitle\abstracts{
We present results from recent investigations of diffractive
vector meson production using the ZEUS detector at HERA. These
consist of measurements of {\rhoz} production in the region of
photon virtuality $0.25 < {\qsq} < 0.85 \; \gevsq$, of
{\rhoz}, $\phi$, and {\jpsi} production for values of the 
momentum transfer to the proton $0.3 < |t| < 4.0 \; \gevsq$, and
of elastic 
{\jpsi} production both in photoproduction and for ${\qsq} > 2 \; \gevsq$.
}
\section{Introduction}
The subject of elastic vector meson production at high energies has attracted
much attention due to its
contributions to the new field of hard diffraction. Such measurements
of {\em exclusive} processes, a rarity at high-energy colliders, 
represent a new opportunity to test calculations within the 
framework of perturbative quantum chromodynamics (pQCD).\cite{review}
Here we present recent measurements of diffractive vector meson production
with the ZEUS experiment
in the interactions of 27.5 GeV positrons with 820 GeV protons at HERA. 
%These
%results update the experimental status of such investigations
%as described in contributions to the 
%XXVIIIth International Conference on High energy Physics in July, 1996 
%(see references below). 
Further details may be found
in papers submitted to the recent DIS '97 Workshop~\cite{dis}. 
We concentrate on
three topics:
\begin{itemize}
\item
elastic production of {\rhoz} mesons in an intermeditate range of photon
virtuality {\qsq}, $0.25 < {\qsq} < 0.85\;\gevsq$.
This analysis is based on measurements recorded using a special-purpose
electromagnetic calorimeter with acceptance at positron scattering angles
between 17 and 35 mrad,
called the beam-pipe calorimeter (BPC),
\item
photoproduction of {\rhoz}, $\phi$, and {\jpsi} mesons 
for high absolute 
values of the squared momentum transferred to the proton, $|t|$, 
$0.3 < |t| < 4.0\;\gevsq$. Since the final-state proton system
(usually dissociated) was not
detected, its transverse momentum was
approximated by the transverse momentum of the final-state
vector meson. The difference is
kinematically limited to a value less than {\qsq}.
Detection of the final-state positron in another special-purpose
calorimeter 
44~m distant from the interaction point (``44-meter tagger''),
limited {\qsq} to values less than 0.01 {\gevsq}, ensuring the validity
of the approximation at that level,
\item
production of {\jpsi} mesons both for untagged photoproduction
($10^{-10} < {\qsq} < 4\;\gevsq$),
and for a sample where the final-state positron was detected in the
central calorimeter (${\qsq} > 2\;{\gevsq}$).
\end{itemize}

Central to these investigations is the question of the validity 
of pQCD calculations,
an issue which has stimulated much interest due in part to the
novelty of applying perturbative methods to
diffractive processes.\cite{many} It has been suggested that
the necessary hard scale may be given by {\qsq}, $t$, or
the vector-meson mass.
Our investigations aim to test these ideas.

\section{Elastic Electroduction of {\rhoz} Mesons}
\begin{figure}[htbp]
\begin{minipage}[t]{0.46\textwidth}
\begin{center}
\epsfig{file=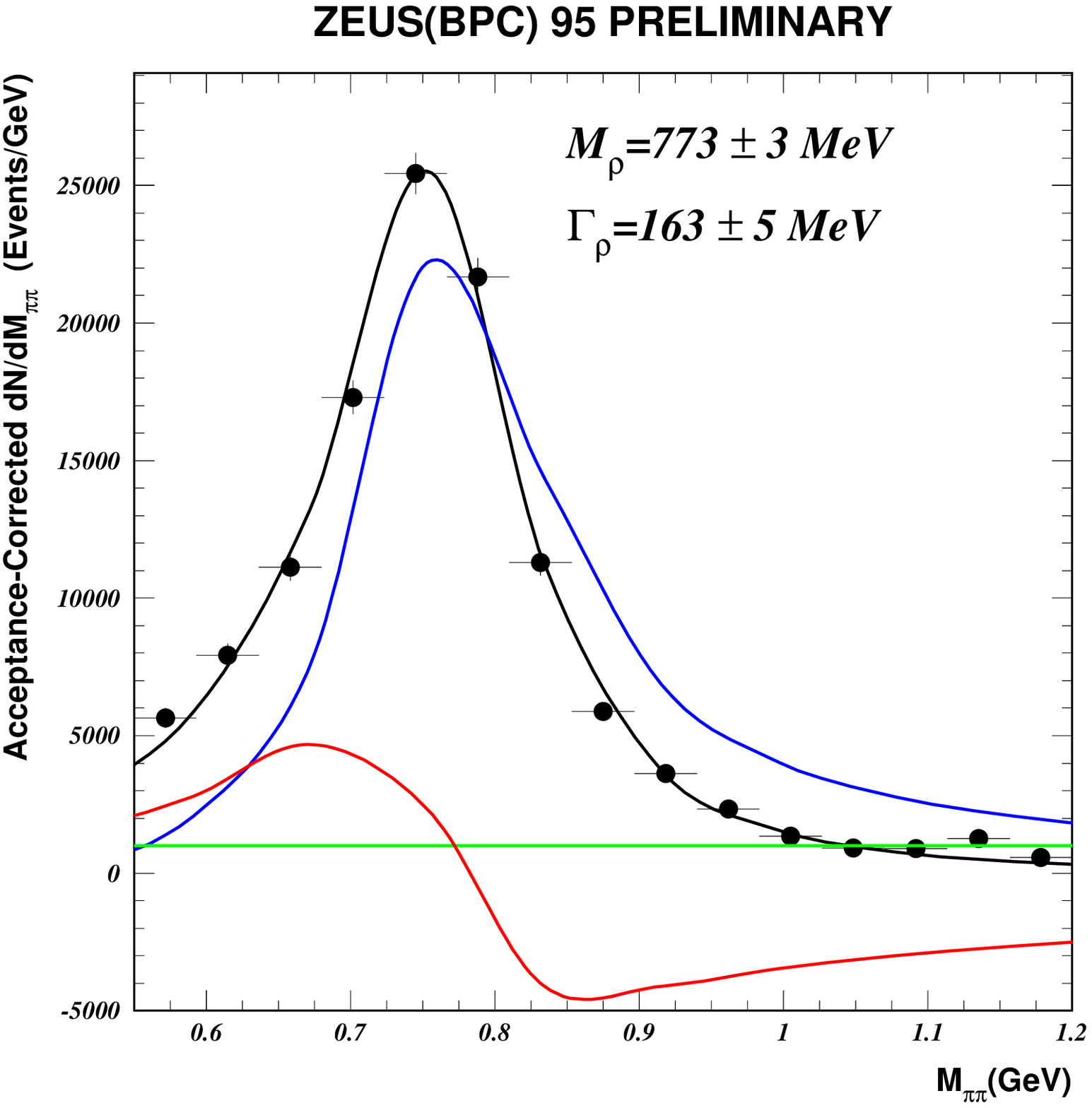,width=0.98\textwidth, bbllx=0pt, bblly=19pt, bburx=461pt, bbury=484pt, clip=}
\vspace*{-5mm}
\end{center}
\caption{
\label{fig:bpc_mass_new}
Acceptance-corrected $\pi^+\pi^-$ invariant mass distribution
for elastic {\rhoz} production at intermediate {\qsq}
\hspace*{\fill}
}
\end{minipage}
\hspace*{0.05\textwidth}
\begin{minipage}[t]{0.46\textwidth}
\begin{center}
\epsfig{file=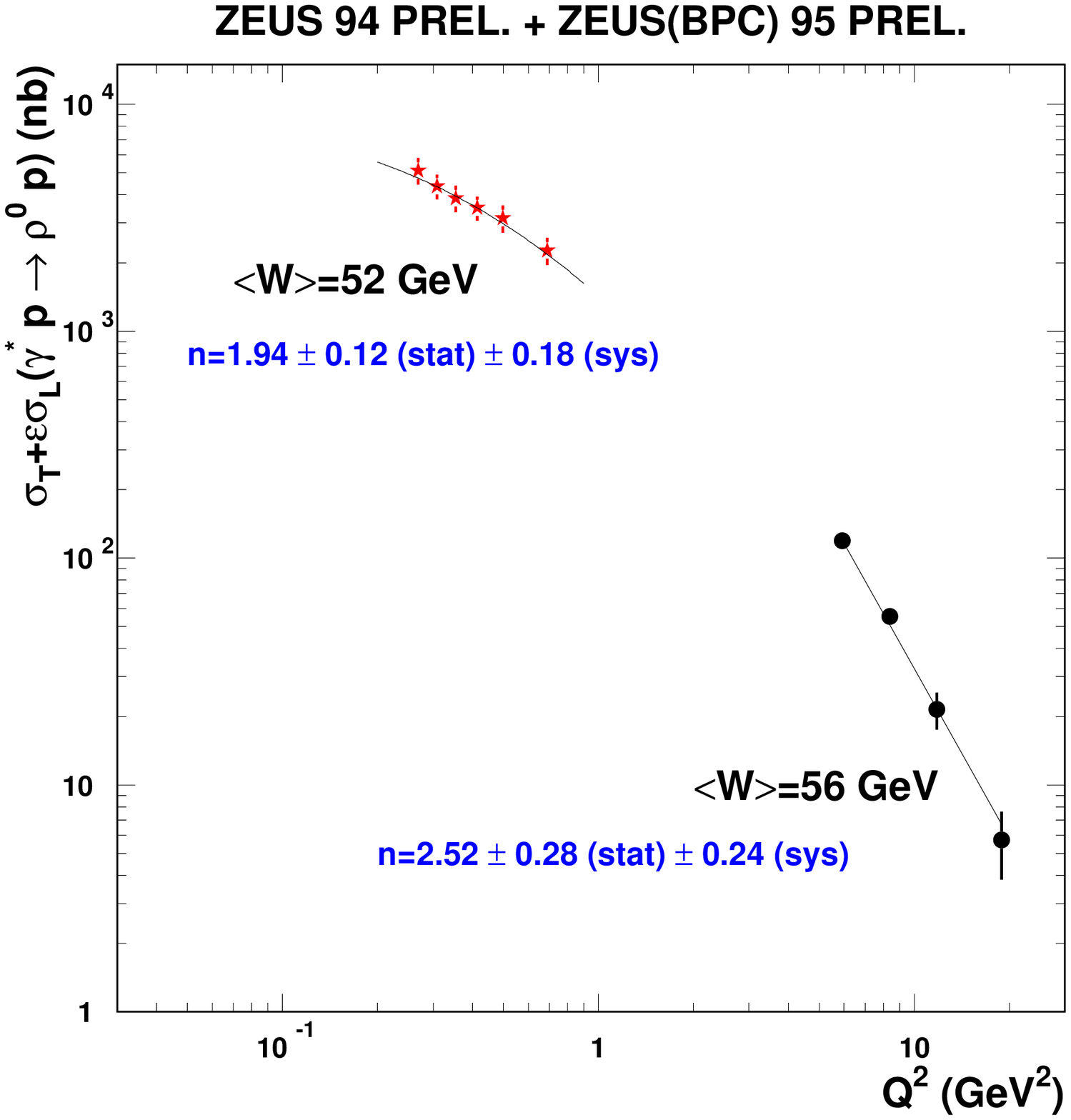,width=0.98\textwidth, bbllx=0pt, bblly=8pt, bburx=512pt, bbury=542pt, clip=}
\vspace*{-5mm}
\end{center}
\caption{
\label{fig:bpc_q2plot}
The {\qsq} dependence of the {\rhoz} elastic cross section  \hspace*{\fill}
}
\end{minipage}
\vspace*{-2mm}
\end{figure}

The exclusive dipion BPC sample 
recorded
in 1995 with an integrated luminosity of 3.8 pb$^{-1}$ 
includes about 6000 events in the {\rhoz} mass region and covers the 
kinematic range 
$0.25 < {\qsq} < 0.85\;\gevsq$, $|t| < 0.6\;\gevsq$,
$20 < W_{\gamma^* {\mathrm p}} < 90\;\gev$, where $W_{\gamma^* {\mathrm p}}$ is the center-of-mass
energy of the virtual-photon-proton system.
The invariant mass spectrum in the {\rhoz} mass region 
is shown in Fig.~\ref{fig:bpc_mass_new}, together with the results
of a fit to the squared coherent sum of a Breit--Wigner resonance term and a nonresonant
background term:
\begin{eqnarray}
\label{eq:dipionspectrum}
\frac{dN}{dM_{\pi \pi}} &=& \left | {A} \frac{\sqrt{M_{\pi \pi}M_{\rho}\Gamma_{\rho}}}{M^2_{\pi \pi}-M^2_{\rho}+iM_{\rho}\Gamma_{\rho}}+ {B} \right |^2\;.
\end{eqnarray}
The dashed line on the mass spectrum indicates 
the sum of the contributions
from the Breit--Wigner resonance term ({\em dotted line}), 
the nonresonant
background term ({\em dash-dotted line}), 
and the interference term ({\em solid line}). 
The resonant cross section averaged over $W_{\gamma^* {\mathrm p}}$ is plotted 
as a function of {\qsq} in Fig.~\ref{fig:bpc_q2plot} 
and compared to the measurements~\cite{pa02_28} at similar
$W_{\gamma^* {\mathrm p}}$ and high {\qsq} in order to investigate the {\qsq} dependence.
The result of a fit to the dependence $({\qsq}+M_{\rho}^2)^{-n}$
yields the result $n=1.94 \pm 0.12({\mathrm stat}) \pm 0.18({\mathrm sys})$,
consistent with pQCD calculations for longitudinal photons. However,
the cross sections measured here are the sum of
contributions from longitudinal and transverse photons $\sigma_{\mathrm L} + \varepsilon \sigma_{\mathrm T}$, where $\varepsilon$ is the ratio of the transverse to longitudinal
photon flux ($0.97 < \varepsilon < 1.00$). This ambiguity
in the comparison to the calculations will be removed when information
on the ratio $R=\sigma_{\mathrm L}/\sigma_{\mathrm T}$ becomes available. 

\begin{figure}[htbp]
\begin{minipage}[t]{0.48\textwidth}
\begin{center}
\epsfig{file=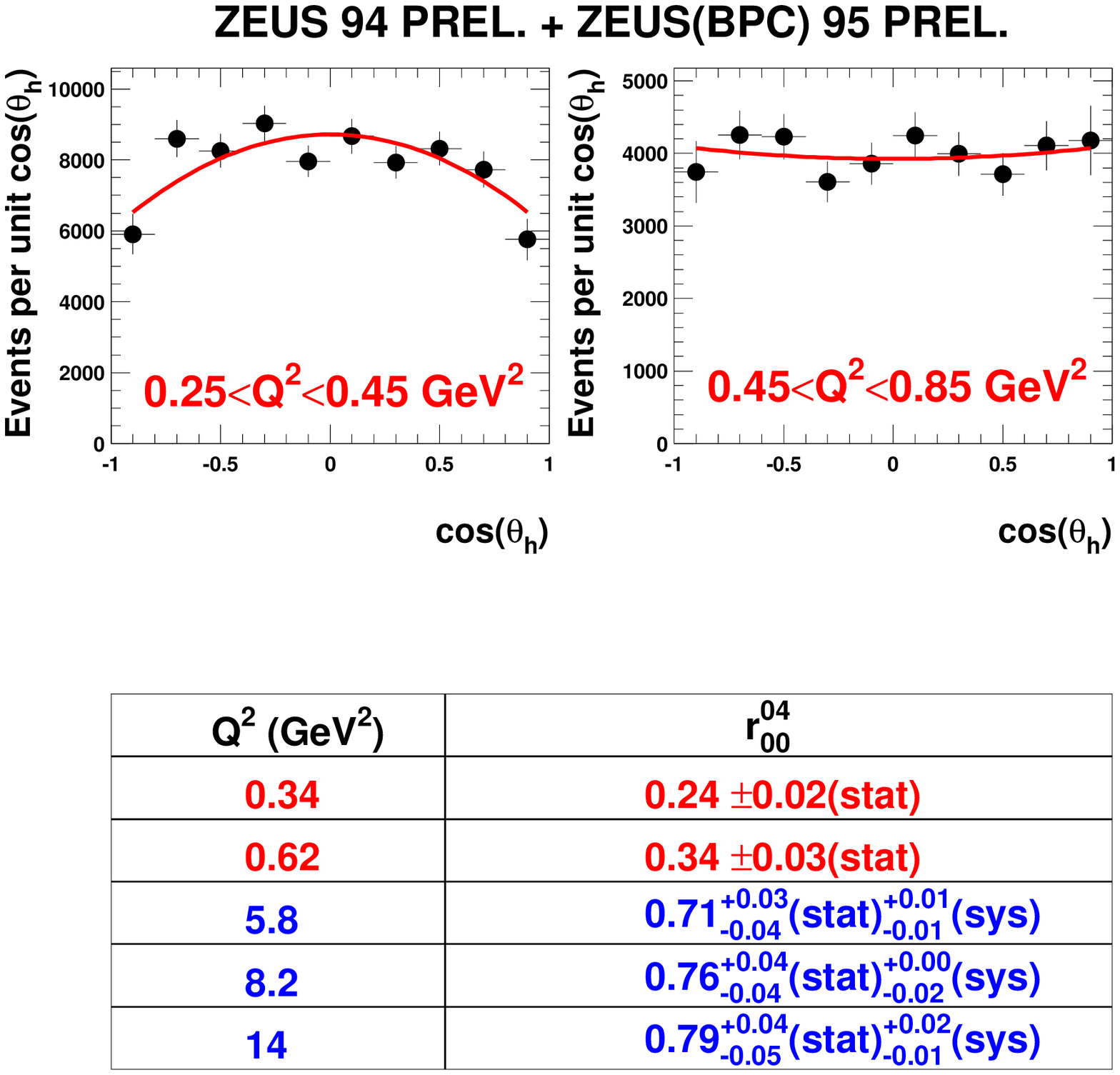,width=0.98\textwidth, bbllx=7pt, bblly=57pt, bburx=512pt, bbury=542pt, clip=}
\vspace*{-3mm}
\end{center}
\caption{
\label{fig:bpc_hel_theta}
Polar decay angle distributions for the BPC {\rhoz} samples and
fit results for $r_{00}^{04}$ 
at intermediate and high {\qsq}  \hspace*{\fill}
}
\end{minipage}
\hspace*{0.02\textwidth}
\begin{minipage}[t]{0.48\textwidth}
\begin{center}
\epsfig{file=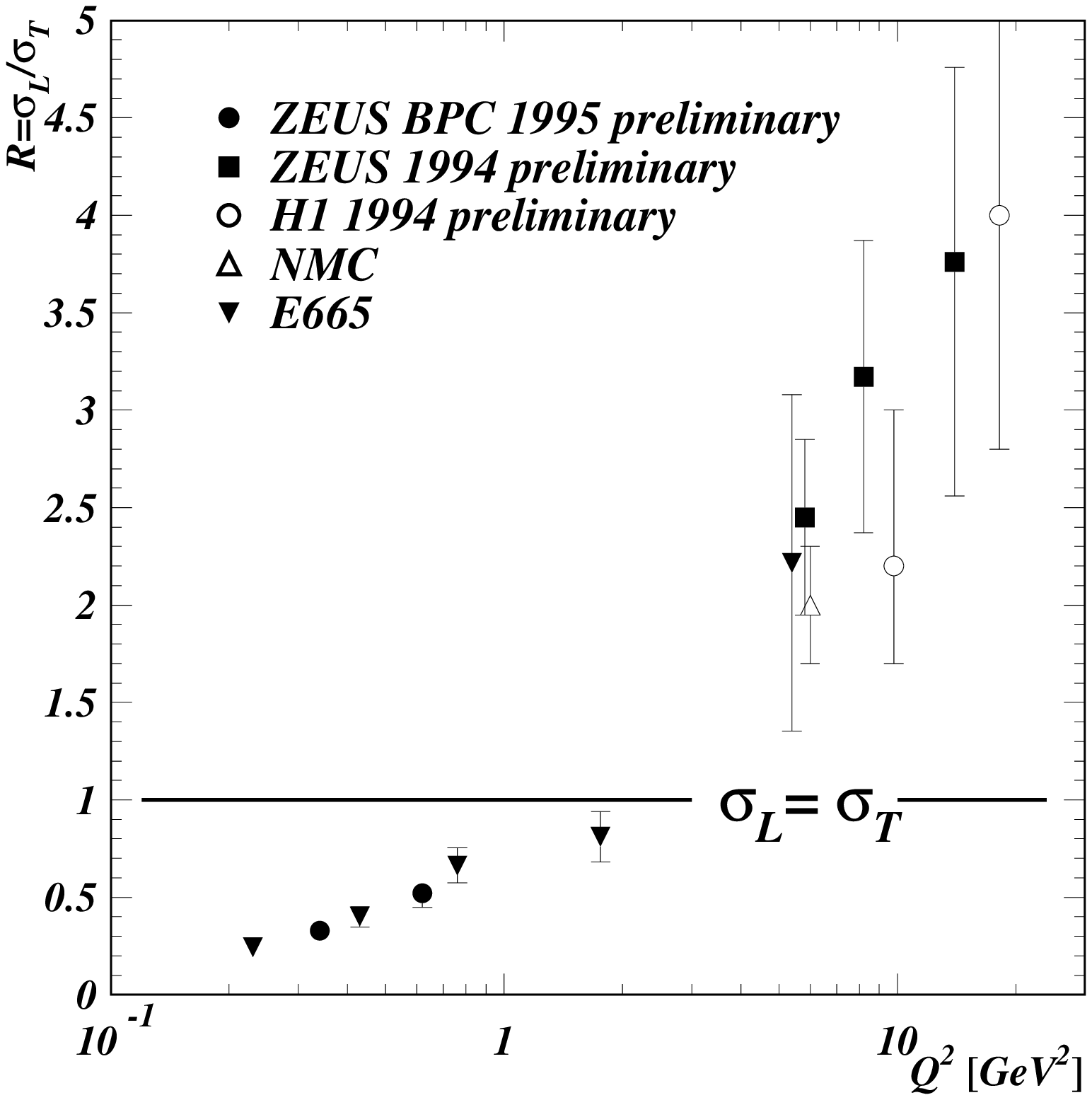,width=0.98\textwidth, bbllx=9pt, bblly=16pt, bburx=455pt, bbury=463pt, clip=}
\vspace*{-3mm}
\end{center}
\caption{
\label{fig:rplot}
Values for $R$ as a function of {\qsq}
derived from the spin-density matrix element $r_{00}^{04}$
under the assumption of SCHC  \hspace*{\fill}
}
\end{minipage}
\vspace*{-2mm}
\end{figure}

A first attempt to measure this ratio is exemplified in 
Figs.~\ref{fig:bpc_hel_theta} and~\ref{fig:rplot}. 
Figure~\ref{fig:bpc_hel_theta} shows the polar angle 
distribution of the $\pi^+$ direction in the {\rhoz} rest system 
with the $Z$ axis defined
as the direction opposite to the final-state proton momentum. This
distribution depends on the spin-density matrix element $r_{00}^{04}$ (the probability that the {\rhoz} is produced with longitudinal
polarization~\cite{np_61_381}):
\begin{eqnarray}
\frac{1}{N} \frac{\D N}{\D (\cos{\thetah})}
&=&\frac{3}{4}\left[1-r_{00}^{04}+(3r_{00}^{04}-1)\cos^2{\thetah}\right].
\end{eqnarray}
The distributions indicate that the BPC data cover a
region of transition to increasingly longitudinal polarization at high {\qsq}.
Assuming
that the $s$-channel helicity conservation (SCHC) observed at low energy
holds at HERA energies, the ratio $R$ is related to this matrix element
as $R = {r_{00}^{04}}/ \varepsilon (1-r_{00}^{04})$. Figure~\ref{fig:rplot}
shows the corresponding $R$ values, comparing them to the ZEUS results~\cite{pa02_28} at high {\qsq}, those from H1,\cite{pa03_48}
and those from fixed target
muoproduction.\cite{nmce665r} The need for this assumption will remain
until full helicity analyses are available.

\section{Elastic {\rhoz}, $\phi$, and {\jpsi} Photoproduction at High $|t|$}
\begin{figure}[htbp]
\begin{minipage}[t]{0.46\textwidth}
\begin{center}
\epsfig{file=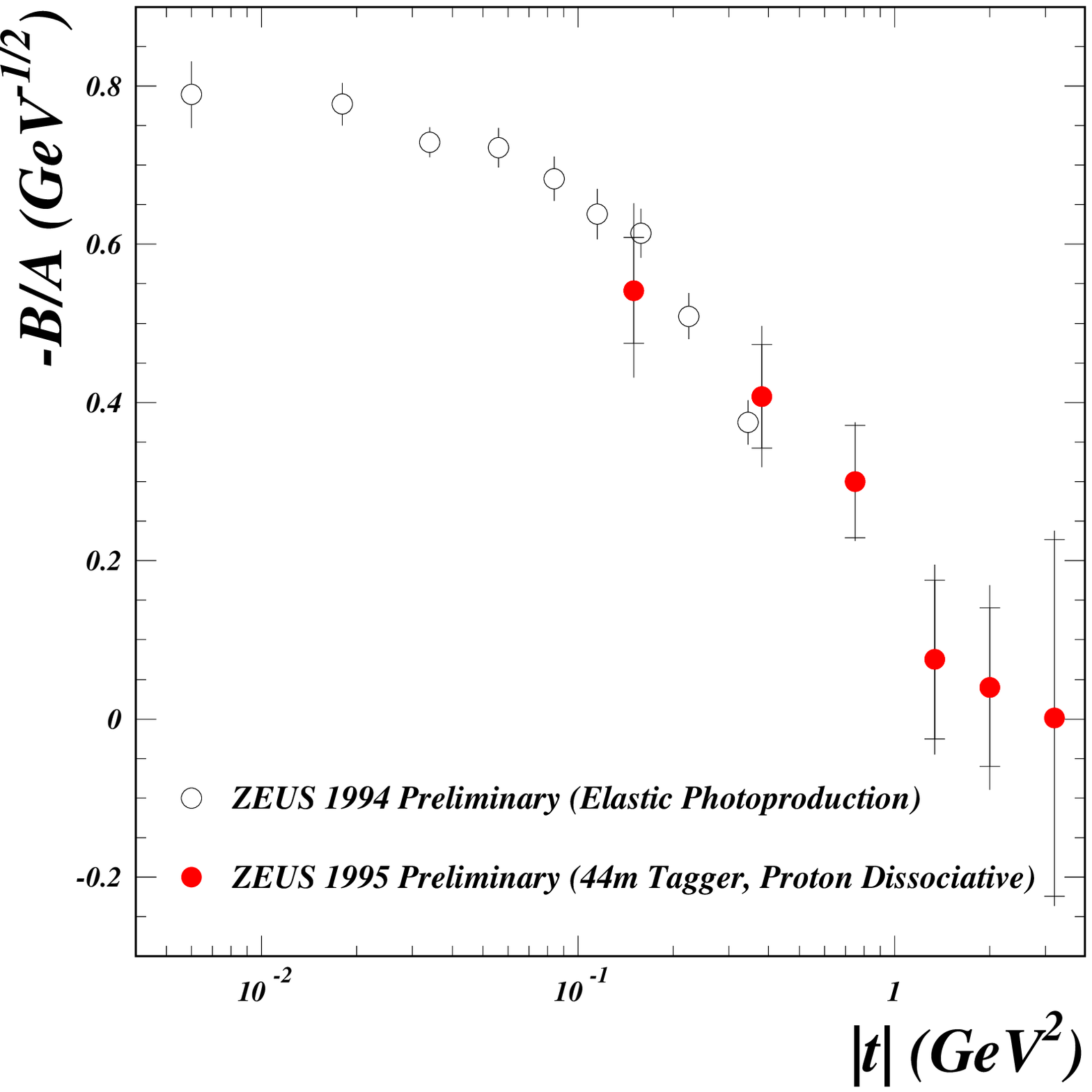,width=0.98\textwidth, bbllx=0pt, bblly=0pt, bburx=456pt, bbury=487pt, clip=}
\vspace*{-5mm}
\end{center}
\caption{
\label{fig:ba_44m}
Ratio of dipion mass spectrum fit parameters, $B/A$, as a function of $t$ 
(eq.~\protect\ref{eq:dipionspectrum}) \hspace*{\fill}
}
\end{minipage}
\hspace*{0.04\textwidth}
\begin{minipage}[t]{0.46\textwidth}
\begin{center}
\epsfig{file=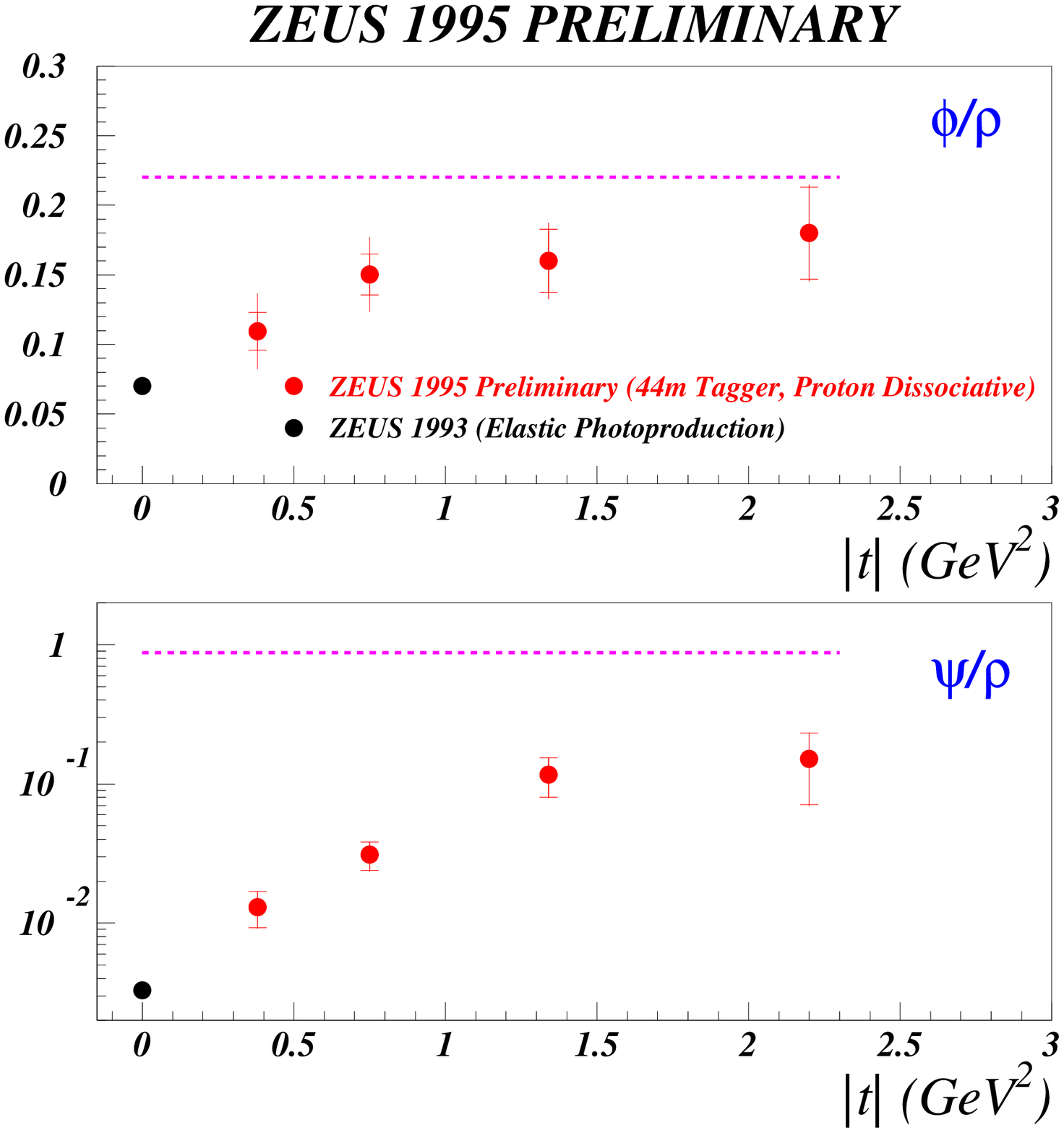,width=0.98\textwidth, bbllx=9pt, bblly=7pt, bburx=514pt, bbury=543pt, clip=}
\vspace*{-5mm}
\end{center}
\caption{
\label{fig:ratios_44m}
Ratios of $\phi$ and {\jpsi} production to {\rhoz} production
as a function of $t$  \hspace*{\fill}
}
\end{minipage}
\vspace*{-2mm}
\end{figure}

The ZEUS collaboration has presented~\cite{pa02_50} a sample
of about 80000 
untagged elastic photoproduced {\rhoz} meson events based on 2.2 pb$^{-1}$ 
recorded in 1994, covering the kinematic range
$10^{-10} < {\qsq} < 4\;\gevsq$, $\langle {\qsq} \rangle \simeq 10^{-2} \gevsq$,
$|t| < 0.5\;\gevsq$, and
$50 < W_{\gamma^* {\mathrm p}} < 100\;\gev$. The statistical accuracy of this sample
allowed the determination of the relative contributions from resonant
and nonresonant dipion production as a function of $t$ (eq.~\ref{eq:dipionspectrum}). The new analysis
of about 30000 events from an integrated luminosity of 2.1 pb$^{-1}$ recorded with the 44-m tagger in 1995 
($10^{-9} < {\qsq} < 0.01\;\gevsq$, $\langle {\qsq} \rangle 
\simeq 10^{-3} \gevsq$,
$|t| < 4.0\;\gevsq$, and
$85 < W_{\gamma^* {\mathrm p}} < 105\;\gev$) allows the extension of this study to
higher $|t|$, as shown in Fig.~\ref{fig:ba_44m}. 
At values of $|t|$ exceeding
1 {\gevsq} the measurements are consistent with no nonresonant background
contribution. The smooth transition and the 
fact that the high $|t|$ data are dominated by proton dissociative processes 
indicate that the pion pair production proceeds independently of the 
proton dissociation.

Clean $\phi$ and {\jpsi} signals have also been observed in the 44-m tagger
sample, allowing the investigation of production ratios relative to the
{\rhoz} as a function of $|t|$, shown in Fig.~\ref{fig:ratios_44m}.
At high $|t|$ the ratios approach values derived from simple
quark-counting rules and flavor-independent production which
result in the expectation for the ratios: {\rhoz} : $\phi$ : {\jpsi} = 9 : 2 : 8.

\section{Photo- and Electroproduction of {\jpsi} Mesons}
\begin{figure}[htbp]
\begin{minipage}[t]{0.47
\textwidth}
\begin{center}
\epsfig{file=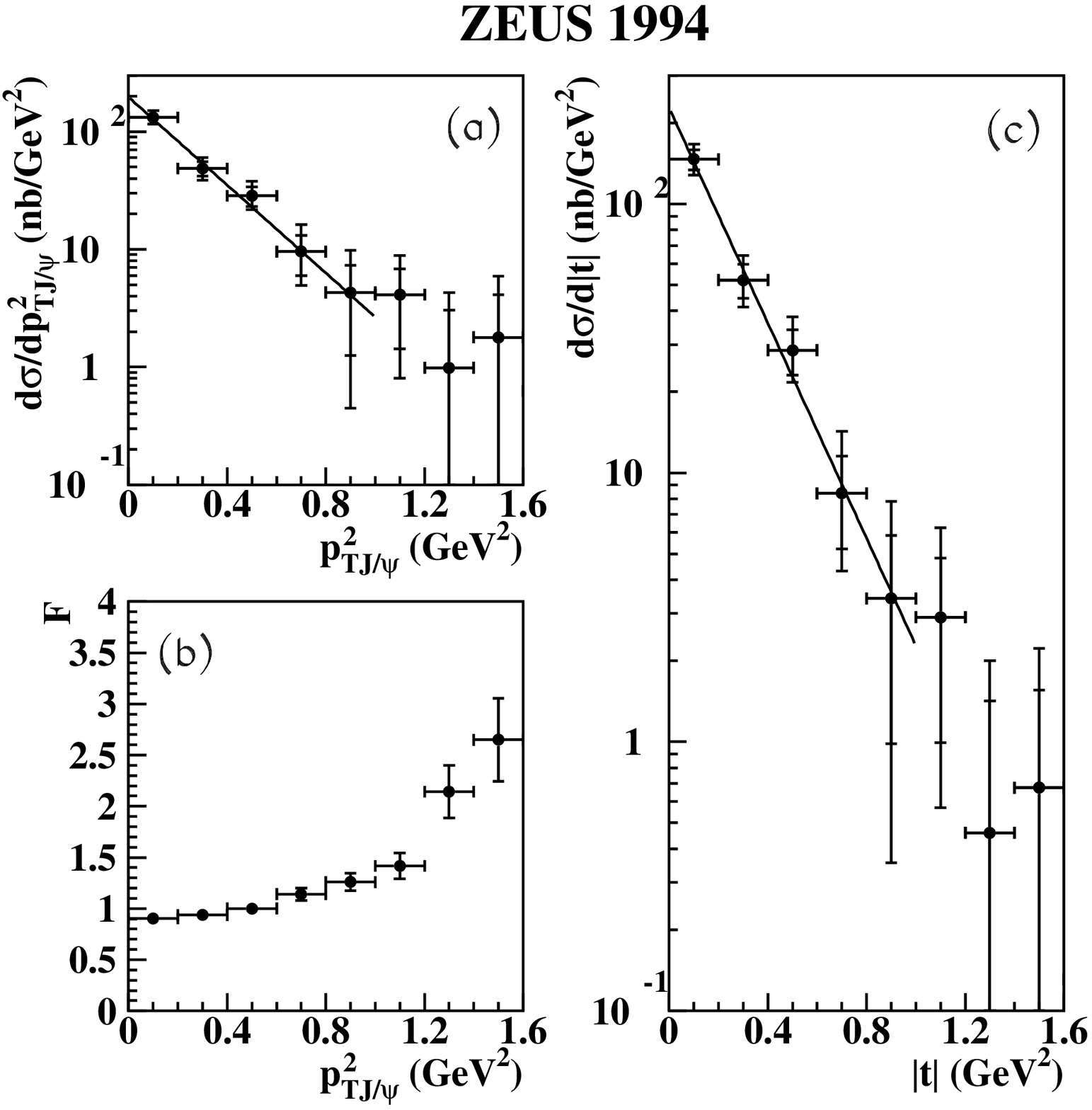,width=0.98\textwidth, bbllx=0pt, bblly=12pt, bburx=527pt, bbury=548pt, clip=}
\vspace*{-5mm}
\end{center}
\caption{
\label{fig:pt2_t}
$p_{\mathrm T}^2$ and $|t|$ dependence of the {\jpsi} elastic photoproduction cross section.   \hspace*{\fill}
}
\end{minipage}
\hspace*{0.035\textwidth}
\begin{minipage}[t]{0.47\textwidth}
\begin{center}
\epsfig{file=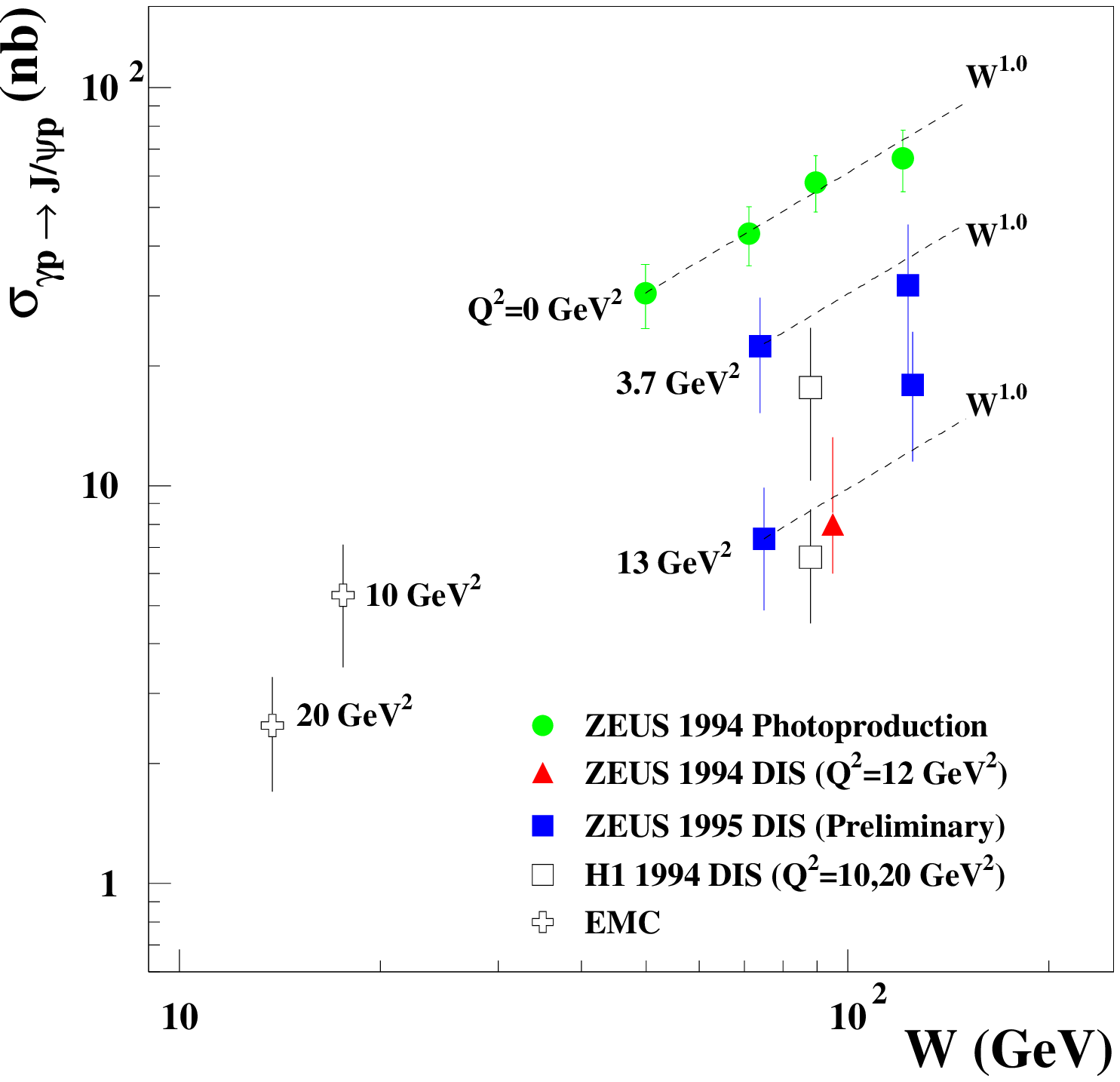,width=0.98\textwidth, bbllx=0pt, bblly=15pt, bburx=428pt, bbury=430pt, clip=}
\vspace*{-5mm}
\end{center}
\caption{
\label{fig:jpsidis_w}
Energy dependence of the {\jpsi} electroproduction 
cross section  \hspace*{\fill}
}
\end{minipage}
\vspace*{-2mm}
\end{figure}

Photoproduction of {\jpsi} mesons for
$10^{-10} < {\qsq} < 4\;\gevsq$, 
$\langle {\qsq} \rangle \simeq 10^{-1} \gevsq$, and
$40 < W_{\gamma^* {\mathrm p}} < 140\;\gev$ has been investigated~\cite{dr_97_060}
via their leptonic decays. An integrated luminosity
of 2.70 pb$^{-1}$ (1.87 pb$^{-1}$) from 1994 
was analyzed for the $e^+e^-$ ($\mu^+\mu^-$) decay mode, yielding
$460 \pm 25$ ($266 \pm 17$) events. 
A fit to the dependence $W_{\gamma^* {\mathrm p}}^\delta$ yields
the result $\delta = 0.92 \pm 0.14({\mathrm stat}) \pm 0.10({\mathrm sys})$,
inconsistent with the exchange of a soft Pomeron alone 
($\delta \simeq 0.22$).\cite{michele} 
%Since {\rhoz} photoproduction shows an energy dependence consistent with
%soft diffraction, these results indicate that the {\jpsi} mass 
%provides the hard scale.
Figure~\ref{fig:pt2_t} shows the extraction of the $|t|$ dependence
in {\jpsi} photoproduction:
({\bf a}) the dependence of the  {\jpsi} photoproduction
cross section on the squared transverse momentum of the
{\jpsi}, $p_{\mathrm T}^2$, ({\bf b}) the factor, $F$, required
to relate the  $p_{\mathrm T}^2$ dependence
to the $|t|$ dependence, derived 
from simulations,
({\bf c}) the $|t|$-dependence of the {\jpsi} elastic photoproduction cross
section. 
The fit to the
function e$^{-b|t|}$ yields 
$b = 4.6 \pm 0.4\, ({\mathrm stat}) ^{+0.4}_{-0.6}\, ({\mathrm sys})\; {\gev}^{-2}$, indicating that the interaction radius is much smaller than that
found~\cite{zfp_73_253} for the  {\rhoz}
($b = 9.8 \pm 0.8\, ({\mathrm stat}) \pm 1.1\, ({\mathrm sys})\; {\gev}^{-2}$).

An analysis of 6 pb$^{-1}$ recorded in 1995 has allowed 
a measurement of the $W_{\gamma^* {\mathrm p}}$ 
dependence of the {\jpsi} electroproduction
cross section for $2 < {\qsq} < 40\;\gevsq$
and $50 < W_{\gamma^* {\mathrm p}} < 150\;\gev$, 
shown in Fig.~\ref{fig:jpsidis_w}. 
A fit to the dilepton mass signal region yields a signal of $101 \pm 13$
events ($e^+e^-$ and $\mu^+\mu^-$ combined).
The $t$-slope is found to be 
$b = 4.5 \pm 0.8\, ({\mathrm stat}) \pm 1.0\, ({\mathrm sys})\; {\gev}^{-2}$,
consistent with the photoproduction result.

%\clearpage
\section{Summary}
Analyses of the {\qsq} dependence of elastic {\rhoz} electroproduction,
of {\rhoz} photoproduction at high values of the momentum transferred to the
target proton, and of the energy dependence in {\jpsi} photoproduction
lend credibility to the idea 
that the variables {\qsq}, $t$, and M$_{\mathrm V}$ each can be used
to define a transition region between soft and hard diffraction.
The ZEUS experiment is sensitive to the transition region in {\em each} of
these variables, allowing detailed, multi-parameter,
studies of the transition from
nonperturbative to perturbative QCD. The studies are statistics-limited
and their precision is expected to improve dramatically when the
more recent data sets have been analyzed.


\section*{References}
\begin{thebibliography}{99.}
\bibitem{review}
For a review of HERA results, see J.A.Crittenden, DESY Report 97-068 (1997), to appear as Nr. 140 in {\em Springer Tracts in Modern Physics}
\relax
\relax
\bibitem{dis}
Talks by L.~Adamczyk, L.~Bellagamba, and T.~Monteiro at the
5th International Workshop on Deep Inelastic Scattering and QCD,
April, 1997 \relax
%, http://www.hep.anl.gov/dis97 \relax
\relax
\bibitem{many}
M.G. Ryskin,
\newblock  Z. Phys. {\bf C 57}  (1993) 89; \relax
\relax
\newblock  M.G. Ryskin {\em et al.}, Preprint HEP--PH/95-11-228 (1995); \relax
\relax
S.J. Brodsky {\em et al.},
\newblock  Phys. Rev. {\bf D 50}  (1994) 3134; \relax
\relax
L. Frankfurt {\em et al.},
\newblock  Phys. Rev. {\bf D 54}  (1996) 3194; \relax
\relax
\newblock  A. Martin {\em et al.}, Preprint HEP--PH/96-09-448 (1996); \relax
\relax
J. Nemchik {\em et al.},
\newblock  Phys. Lett. {\bf 374}  (1996) 199; \relax
\relax
D.Yu. Ivanov,
\newblock  Phys. Rev. {\bf D 53}  (1996) 3564; \relax
\relax
I.F. Ginzburg and D.Yu. Ivanov,
\newblock  Phys. Rev. {\bf D 54}  (1996) 5523; \relax
\relax
J.R. Forshaw {\em et al.},
\newblock  Z. Phys. {\bf C 68}  (1995) 137 \relax
\relax
\bibitem{pa02_28}
The ZEUS Collaboration,
\newblock  PA02--028, XXVIII International Conference on High Energy Physics,
  Warsaw, Poland, July 25--31, 1996\relax
\relax
\bibitem{np_61_381}
K. Schilling and G. Wolf,
\newblock  Nucl. Phys. {\bf B 61}  (1973) 381\relax
\relax
\bibitem{pa03_48}
The H1 Collaboration,
\newblock  PA03--048, XXVIII International Conference on High Energy Physics,
  Warsaw, Poland, July 25--31, 1996\relax
\relax
\bibitem{nmce665r}
The NMC Collaboration,
\newblock  Nucl. Phys. {\bf B 429}  (1994) 503; \relax
\relax
The E665 Collaboration,
\newblock  Z. Phys. {\bf C 72}  (1997) 237\relax
\relax
\bibitem{pa02_50}
The ZEUS Collaboration,
\newblock  PA02--050, XXVIII International Conference on High Energy Physics,
  Warsaw, Poland, July 25--31, 1996\relax
\relax
\bibitem{dr_97_060}
The ZEUS Collaboration,
\newblock  DESY Report 97-060 (1997)\relax
\relax
\bibitem{michele}
M.~Arneodo,
\newblock  these Proceedings\relax
\relax
\bibitem{zfp_73_253}
The ZEUS Collaboration,
\newblock  Z. Phys. {\bf C 73}  (1997) 253\relax
\relax

\end{thebibliography}
\end{document}